\documentclass[nofootinbib,aps,12pt]{revtex4-1}
\usepackage{graphicx}
\usepackage{cancel}
\usepackage{textcomp}
\usepackage{amsmath}
\usepackage{bm}
\usepackage{epsfig}
\usepackage{color}
\usepackage{epstopdf}
\usepackage{dcolumn}
\def\beq{\begin{equation}}
\def\eeq{\end{equation}}

\def\be{\begin{equation}}
\def\ee{\end{equation}}
\def\bea{\begin{eqnarray}}
\def\eea{\end{eqnarray}}

\begin{document}
\title{ LHC Constraints on NLSP Gluino and Dark Matter Neutralino in Yukawa Unified Models
\vspace{4mm}
}
\author{M. Adeel Ajaib\footnote{email: adeel@udel.edu}}
\author{Tong Li\footnote{email: tli@udel.edu, corresponding author}}
\author{Qaisar Shafi\footnote{email: shafi@bartol.udel.edu}}
\address{
Bartol Research Institute, Department of Physics and Astronomy,
University of Delaware, Newark, Delaware 19716, USA
\vspace{3mm}
}

\begin{abstract}
The ATLAS experiment has recently presented its search results for final states containing jets and/or b-jet(s) and missing transverse momentum, corresponding to an integrated luminosity of 165 pb$^{-1}$. We employ this data to constrain a class of  supersymmetric $SU(4)_c\times SU(2)_L\times SU(2)_R$ models with $t-b-\tau$ Yukawa unification, in which the gluino is the next to lightest supersymmetric particle (NLSP). The NLSP gluino is slightly ($\sim$10-30\%) heavier than the the LSP dark matter neutralino, and it primarily decays into the latter and a quark-antiquark pair or gluon. We find that NLSP gluino masses below $\sim$ 300 GeV are excluded by the ATLAS data. For LSP neutralino mass $\sim 200-300$ GeV and $\mu>0$, where $\mu$ is the coefficient of the MSSM Higgs bilinear term, the LHC constraints in some cases on the spin-dependent (spin-independent) neutralino-nucleon cross section are  significantly more stringent than the expected bounds from IceCube DeepCore (Xenon 1T/SuperCDMS). For $\mu<0$, this also holds for the spin-dependent cross sections.
\end{abstract}
 \maketitle

\section{I\lowercase{ntroduction}}
Low scale supersymmetry, augmented by an unbroken $Z_2$ matter (R-) parity, largely overcomes the gauge hierarchy problem encountered in the Standard Model (SM) and also provides a compelling cold dark matter candidate. In the mSUGRA/constrained minimal supersymmetric model (CMSSM)~\cite{Arnowitt:2006bb} , as well as in many other realistic models, the lightest neutralino (LSP) is stable~\cite{lsp} with a relic density that is compatible with the WMAP dark matter measurements~\cite{wmap}. However, the small annihilation cross section of a pure bino LSP with mass of around 100 GeV does not permit one to easily reproduce the required relic dark matter abundance~\cite{bino}.

An interesting scenario which enhances the bino annihilation cross section is bino-gluino co-annihilation. In this case the bino and the relevant NLSP gluino (where NLSP stands for next to lightest supersymmetric particle) are sufficiently close together in mass, such that the ensuing co-annihilation processes in the early universe allow one to reproduce the desired bino relic density. This scenario is not possible in the CMSSM, but it has been implemented in models with non-universal gaugino masses~\cite{coann3}, and in a class of (third family) Yukawa unified models~\cite{4221,4222,gy}. The collider signatures of the gluino co-annihilation scenario have recently been discussed in Refs.~\cite{gluino,gluino1}.

The ATLAS and CMS experiments at $\sqrt{s}=7$ TeV LHC have previously presented their search results for low-energy supersymmetry corresponding to an integrated luminosity of 35 pb$^{-1}$~\cite{atlas1,atlas2}, which was recently updated by ATLAS to 165 pb$^{-1}$ ~\cite{atlasnew}. The successful launch of the LHC and a flurry of supersymmetry related papers from the ATLAS and CMS collaborations provides a strong impetus to explore regions of the MSSM parameter space not covered by the minimal version (CMSSM/mSUGRA). In this paper we study the constraints and implications of recent LHC data on some well-motivated NLSP gluino models induced by gaugino mass non-universality and $t-b-\tau$ Yukawa coupling unification imposed at $M_{GUT}$. The underlying symmetry group we consider is $SU(4)_c\times SU(2)_L\times SU(2)_R$~\cite{patisalam}. With the NLSP gluino and LSP neutralino having nearly degenerate masses, the chargino as well as leptons are absent in the gluino cascade decay. Also, the jets and missing energy from NLSP gluino decay are much softer due to the small mass difference between the NLSP and LSP. Thus, the conventional search strategy with same-sign chargino signature does not work here, and the usual requirement of large $p_T$ jet and missing transverse momentum makes the event selection harder to implement. The LHC constraints on the NLSP gluino mass turn out to be significantly less restrictive than the recent 1 TeV or so mass bound on the gluino mass which, among other things, assume an essentially `massless' neutralino.

The paper is organized as follows. In section II we briefly summarize the NLSP gluino scenario with $t-b-\tau$ Yukawa unification and neutralino (essentially bino) dark matter. We also discuss the NLSP gluino decay modes and outline the selection cuts employed by the ATLAS collaboration. The results of two classes of NLSP gluino models constrained by the LHC data 
are presented together with a few benchmark points in section III. Our conclusions are summarized in 
section IV.
 
\section{NLSP G\lowercase{luino}  \lowercase{and} ATLAS S\lowercase{election} C\lowercase{uts}}
\label{sec2}
As mentioned earlier, the gluino-bino co-annihilation scenario requires the gluino to be NLSP in the sparticle spectrum, and to be nearly degenerate in mass with the bino LSP. The mass difference between the two should be~\cite{coann3}
\begin{eqnarray}
{M_{\tilde{g}}-M_{\tilde{\chi}_1^0}\over M_{\tilde{\chi}_1^0}}\lesssim 20\%.
\label{coann}
\end{eqnarray} 
In the framework of minimal supergravity, this feature clearly requires non-universal gaugino masses at $M_{GUT}$. In particular, a partial unified model given by $SU(4)_c\times SU(2)_L\times SU(2)_R$ (4-2-2) group structure provides solutions to this scenario. Non-universal asymptotic gaugino masses are naturally accommodated in the supersymmetric 4-2-2 model and have recently been investigated in Refs.~\cite{4221,4222,gy}. With the SM hypercharge in 4-2-2 given by $Y=\sqrt{2/5}(B-L)+\sqrt{3/5}I_{3R}$, one has the asymptotic relation between the three gaugino masses
\begin{eqnarray}
M_1&=&{3\over 5} M_2+{2\over 5}M_3,
\end{eqnarray}
where $M_1$, $M_2$ and $M_3$ denote the asymptotic gaugino masses of $U(1)_Y$, $SU(2)_L\times SU(2)_R$ and $SU(3)_c$ respectively. Assuming that charged fermions of the third family acquires mass solely from a single (1,2,2) representation in 4-2-2 leads to the Yukawa unification condition at $M_{GUT}$ \cite{big-422}
\begin{eqnarray}
y_t=y_b=y_\tau\equiv y_{Dirac}.
\end{eqnarray}
It has been shown that $t-b-\tau$ Yukawa unification can yield relatively light gluinos ($\leq$ 1 TeV)~\cite{4221,Baer}.

\begin{table}[tb]
\begin{center}
\begin{tabular}[t]{|c|c|c|c|c|}
  \hline
 & S1 & S2 & S3 & b\\
  \hline
  Number of jets & $\geq 2$ & $\geq 3$ & $\geq 4$ & $\geq 3$\\
  \hline
  Number of $b$-jets & 0 & 0 & 0 & $\geq 1$\\
  \hline
 Leading jet $p_T$ (GeV) & $>130$ & $>130$ & $>130$ & $>120$\\
  \hline 
  Other jets $p_T$ (GeV) & $>40$ & $>40$ & $>40$ & $>30$ \\
  \hline
  $\Delta \phi(\vec{p}_T^{{\rm miss}}, j_{1,2,3})$ & $>0.4$ & $>0.4$ & $>0.4$ & $>0.4$ \\
  \hline
  $m_{eff}$ (GeV) & $>1000$ & $>1000$ & $>1000$ & $>600$\\
  \hline
  $\cancel{E}_T$ (GeV) & $>130$ & $>130$ & $>130$ & $>100$\\
  \hline
  $\cancel{E}_T/m_{eff}$ & $>0.3$ & $>0.25$  & $>0.25$  & $>0.2$ \\
  \hline 
  ATLAS $\sigma_{{\rm exp}}$ (pb) & $35$ & $30$ & $35$ & $0.32$\\
  \hline
\end{tabular}
\end{center}
\caption{Summary of selection cuts and 95$\%$ C.L. upper limits on effective cross section for non-SM processes for signal region S1, S2, S3 with 165 pb$^{-1}$ luminosity, and region b with 35 pb$^{-1}$ luminosity, following ATLAS data analyses~\cite{atlas2,atlasnew}.}
\label{cuts}
\end{table}  


In order to implement radiative electroweak breaking consistent with Yukawa unification, the soft mass terms of the two Higgs doublets must be non-universal at $M_{GUT}$, such that the fundamental parameters in this class of models are
\begin{eqnarray}
m_0,m_{H_u},m_{H_d},M_2,M_3,A_0,\tan\beta,{\rm sign}(\mu).
\label{params}
\end{eqnarray}
Here $m_0$ is the universal soft mass of sfermions, $A_0$ is the universal trilinear scalar coupling, $\tan\beta$ is the ratio of the vacuum expectation values (VEVs) of the two MSSM Higgs doublets, and $\mu$ is the MSSM bilinear Higgs mass parameter. The software package ISAJET 7.80~\cite{isajet} was employed in Refs.~\cite{4221,4222,gy} to scan over the relevant parameter space, including renormalization group evolution of gauge and Yukawa couplings and all soft parameters, as well as the computation of the physical masses of all particles. 
A large number of relevant phenomenological constraints such as $BR(B_s\to \mu^+\mu^-)$~\cite{bs}, $BR(b\to s\gamma)$~\cite{bsr}, $BR(B_u\to \tau\nu)$~\cite{bsr}, $\Delta(g-2)_\mu$~\cite{g-2}, WMAP relic density~\cite{wmap}, LEP II bound on the lightest Higgs and all the sparticle mass bounds~\cite{mass} are also implemented.  
The degree Yukawa of unification is quantified by the parameter $R$ [22, 6, 7]
\begin{eqnarray}
R\equiv {{\rm max}(y_t,y_b,y_\tau)\over {\rm min}(y_t,y_b,y_\tau)} \ .
\end{eqnarray}
We shall require that $R \leq 1$, so that  $t-b-\tau$ Yukawa unification holds at 10\% level or better. Note that the NLSP gluino scenario with nearly-degenerate gluino and bino masses can be realized in 4-2-2 models for both $\mu>0$ and $\mu<0$~\cite{4221,4222,gy}.

Because of the mass degenerate feature in Eq.~(\ref{coann}), the NLSP gluino essentially decays into colored SM particles such as the gluon octet or a quark-antiquark pair, and the color singlet LSP $\tilde{\chi}_1^0$:
\begin{eqnarray}
\tilde{g}\to q\bar{q}\tilde{\chi}_1^0, b\bar{b}\tilde{\chi}_1^0, g\tilde{\chi}_1^0,
\end{eqnarray}
where $q(\bar{q})$ denotes the first two generation quark (antiquark). The three-body decay $\tilde{g}\to q\bar{q}\tilde{\chi}_1^0, b\bar{b}\tilde{\chi}_1^0$ proceeds through an off-shell squark exchange, while the two-body decay $\tilde{g}\to g\tilde{\chi}_1^0$ involves a loop diagram containing squarks and quarks. The partial widths of these two decay channels are given by~\cite{loop1,loop}
\begin{eqnarray}
\Gamma(\tilde{g}\to g \tilde{\chi}^0_{1})&=&
{(M_{\tilde{g}}^2-M_{\tilde{\chi}_1^0}^2)^3\over 2\pi M_{\tilde{g}}^3}[{g_3^2g_1\over 128\pi^2}(M_{\tilde{g}}-M_{\tilde{B}})\sum_qQ_q({1\over M_{\tilde{q}_L}^2}-{1\over M_{\tilde{q}_R}^2})N_{1B}\nonumber \\
&+&{g_3^2y_t^2\over 32\sqrt{2}\pi^2\sin\beta}
({1\over M_{\tilde{q}_L}^2}+{1\over M_{\tilde{u}_R}^2})N_{1H_u}v(1+{\rm ln}{m_t^2\over M_{\tilde{g}}^2})]^2,
\label{loopdecay}\\
\Gamma(\tilde{g}\to q\bar{q} \tilde{\chi}^0_{1})&=&
{M_{\tilde{g}}^5\over 768\pi^3}[({g_3g_1\over 6 M_{\tilde{q}_L}^2}N_{1B}+{g_3g_2\over 2 M_{\tilde{q}_L}^2}N_{1W})^2+({2g_3g_1\over 3M_{\tilde{u}_R}^2}N_{1B})^2\nonumber \\
&+&({g_3g_1\over 6 M_{\tilde{q}_L}^2}N_{1B}-{g_3g_1\over 2 M_{\tilde{q}_L}^2}N_{1W})^2+({g_3g_1\over 3M_{\tilde{d}_R}^2}N_{1B})^2]f({M_{\tilde{\chi}^0_{1}}\over M_{\tilde{g}}}) \ (q=u,d),\\
f(x)&=&1+2x-8x^2+18x^3-18x^5+8x^6-2x^7-x^8\nonumber \\
&-&12x^4{\rm ln}x^2+12x^3(1+x^2){\rm ln}x^2.
\label{treedecay}
\end{eqnarray}

\begin{figure}[tb]
\begin{center}
\includegraphics[scale=1,width=8cm]{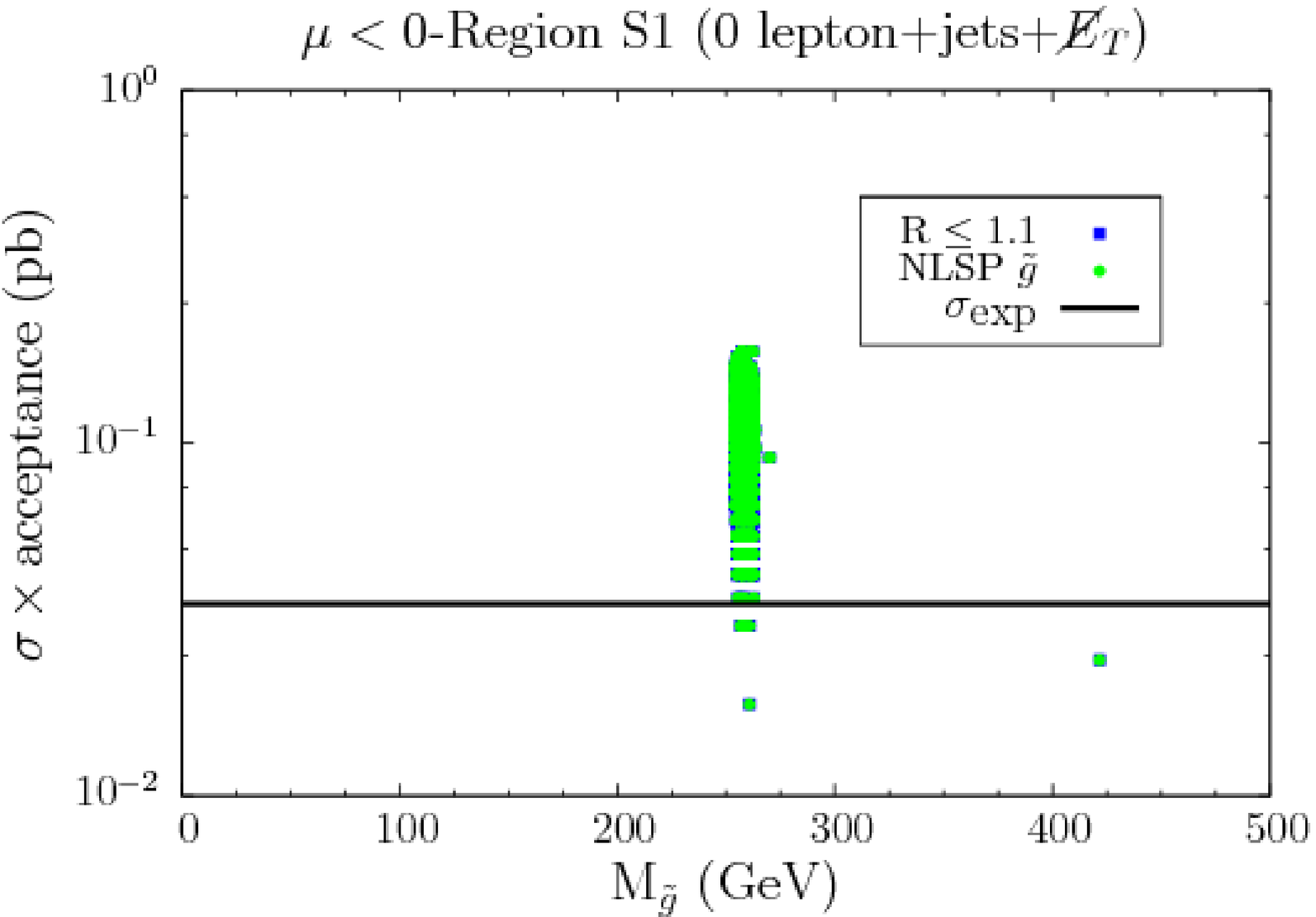}
\includegraphics[scale=1,width=8cm]{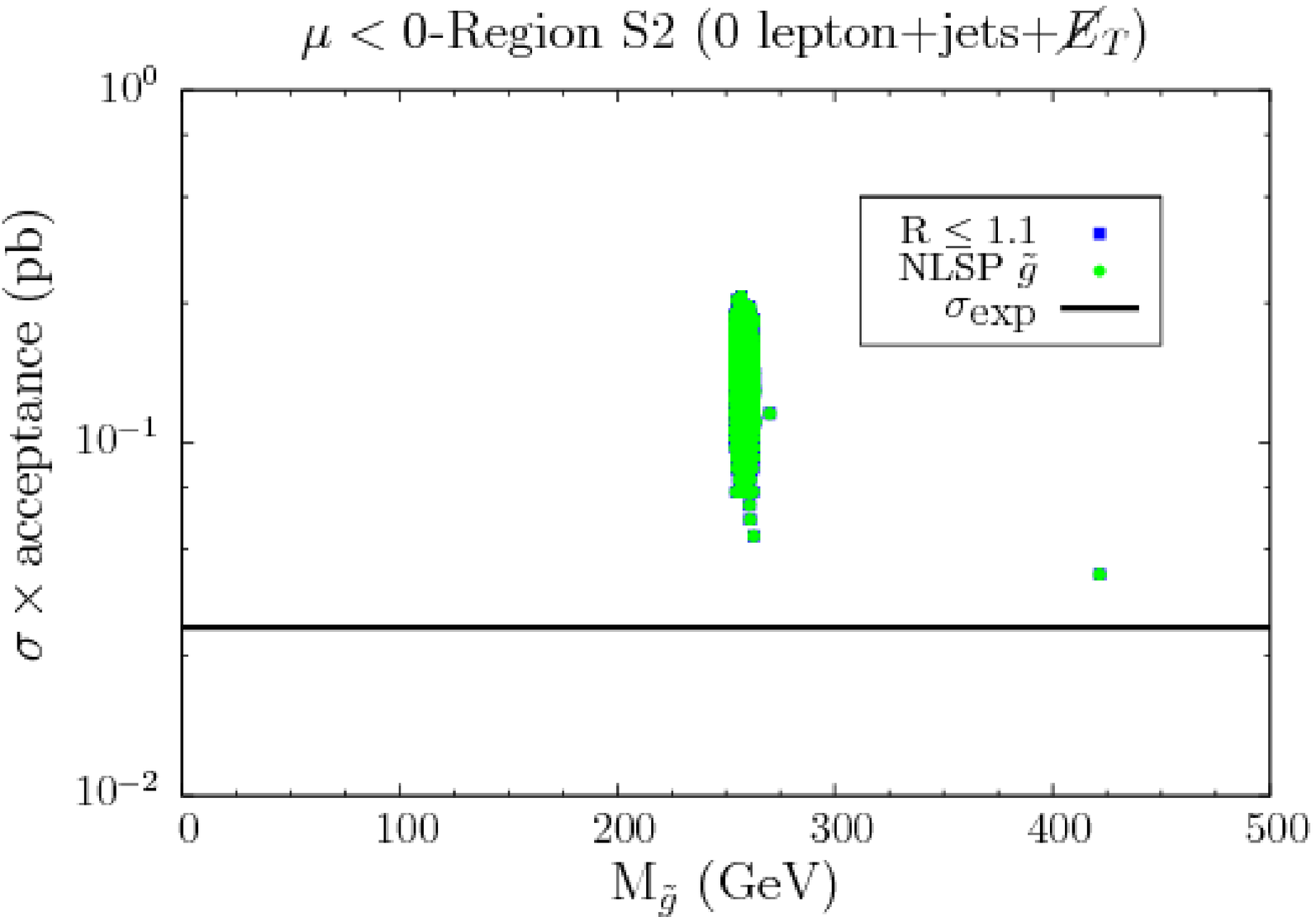}\\
\includegraphics[scale=1,width=8cm]{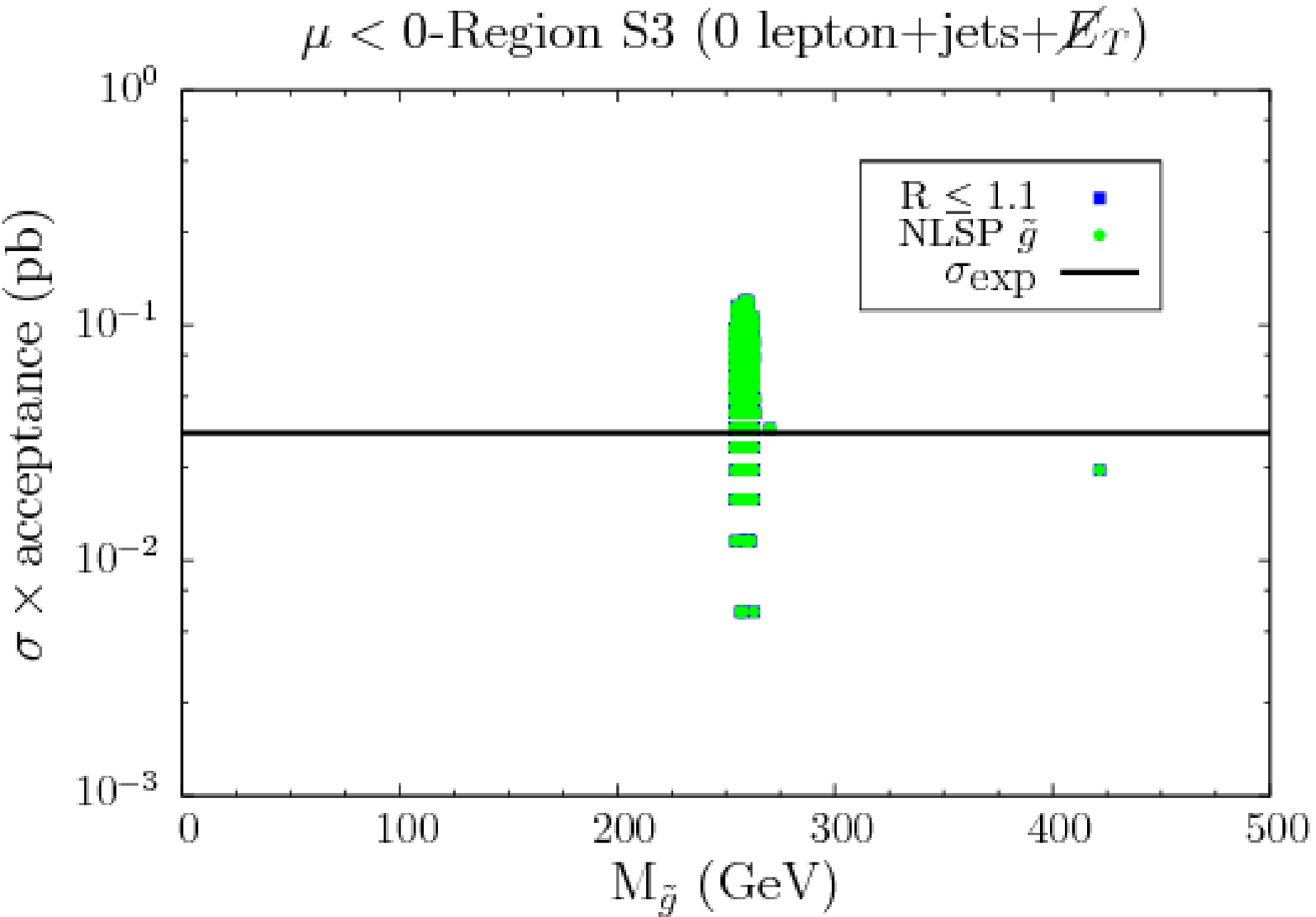}
\includegraphics[scale=1,width=8cm]{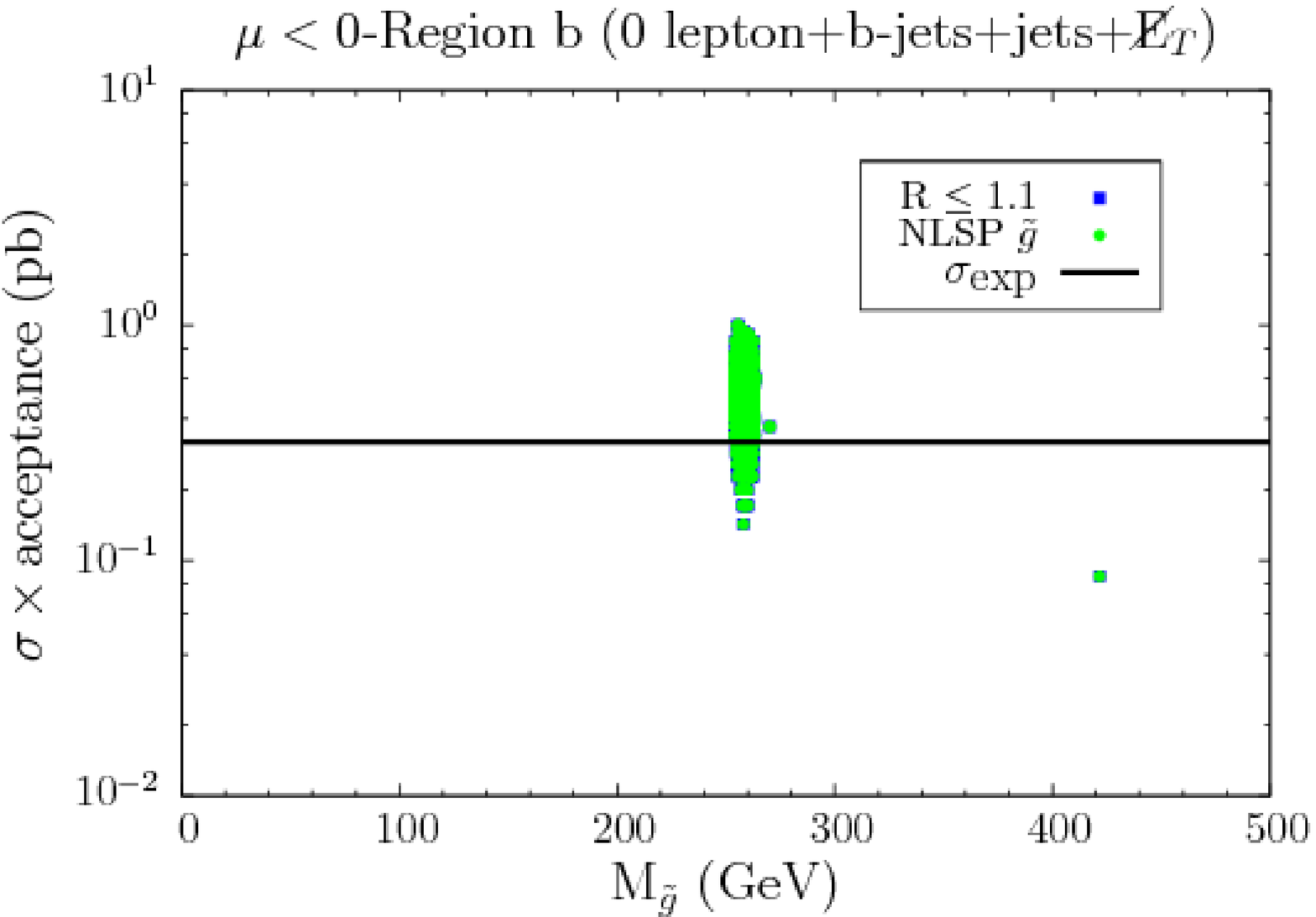}
\end{center}
\caption{$\sigma\times$acceptance vs. $M_{\tilde{g}}$ with horizontal line as the 95$\%$ C.L. upper limits on effective cross section for non-SM processes for signal region S1 (top left), S2 (top right), S3 (bottom left), b (bottom right) in the 4-2-2 framework with $\mu<0$. Blue regions correspond to models with Yukawa unification ($R\leq 1.1$). NLSP gluino models form a subset of these and are represented by green points.} 
\label{nmusigma}
\end{figure}
Here $N_{1B}$, $N_{1W}$ and $N_{1H_u}$ respectively denote the bino, wino and Higgsino components of the LSP neutralino $\tilde{\chi}_1^0$. Generally, the three-body decays will be suppressed if the scalar masses are too large, or by phase space if the mass difference between $\tilde{g}$ and $\tilde{\chi}_1^0$ ($\Delta M\equiv M_{\tilde{g}}-M_{\tilde{\chi}_1^0}$) is too small. 
Assuming $M_{\tilde{\chi}_1^0}\sim \mathcal{O}(250)$, together with the co-annihilation requirement in Eq.~(\ref{coann}), one has the mass difference $\Delta M\simeq 50$ GeV. Also, for large $\tan\beta$, a large bottom Yukawa $y_b$ naturally leaves the bottom squark (sbottom) to be the lightest squark, of $\mathcal{O}$(TeV). With $\Delta M\simeq 50$ GeV and $\mathcal{O}$(TeV) sbottom, $\tilde{g}\to b\bar{b}\tilde{\chi}_1^0$ decay often dominates. One can see this feature from Fig.~1 in Ref.~\cite{gluino1}, which shows the dependence of the gluino decay branching fraction in the $M_{\tilde{g}}-M_{\tilde{b}_1}$ plane for the 4-2-2 model with $\mu<0$. The NLSP gluino decay is therefore sensitive to signals with multi-jets plus missing energy, and relatively more to final states with $b$-jets.

The ATLAS and CMS collaborations have previously reported data in terms of events containing large missing transverse momentum and jets (with or without b-jets) in $\sqrt{s}=7$ TeV proton-proton collisions, corresponding to an integrated luminosity of 35 pb$^{-1}$. No excess above the Standard Model (SM) background expectation was observed~\cite{atlas1,atlas2}. More recently, the ATLAS experiment has considered multi-jets plus missing energy events, with an integrated luminosity of 165 pb$^{-1}$~\cite{atlasnew}. With more strict selection cuts, new lower bounds on non-SM cross-sections that are 30 times more stringent than from the 2010 data have been obtained. This analyses can also be employed, as we show here, to find useful constraints on NLSP gluino models with nearly degenerate gluino and LSP neutralino masses. 

Note that gluino masses below 725 GeV are excluded at the 95\% confidence level in simplified models containing only squarks of the first two generations, gluino and ``massless'' LSP neutralino~\cite{atlasnew}. In this case, with the gluino and squarks much heavier than LSP neutralino, the large mass difference results in highly energetic jets and large missing energy. With nearly-degenerate NLSP gluino and LSP neutralino, however, the jets from gluino decay and missing energy are softer, and fewer events with the same gluino mass would pass the same selection cuts. We therefore expect that the upper limit on the excluded gluino mass for degenerate NLSP gluino and LSP neutralino scenarios would be correspondingly lower.

\begin{figure}[tb]
\begin{center}
\includegraphics[scale=1,width=10cm]{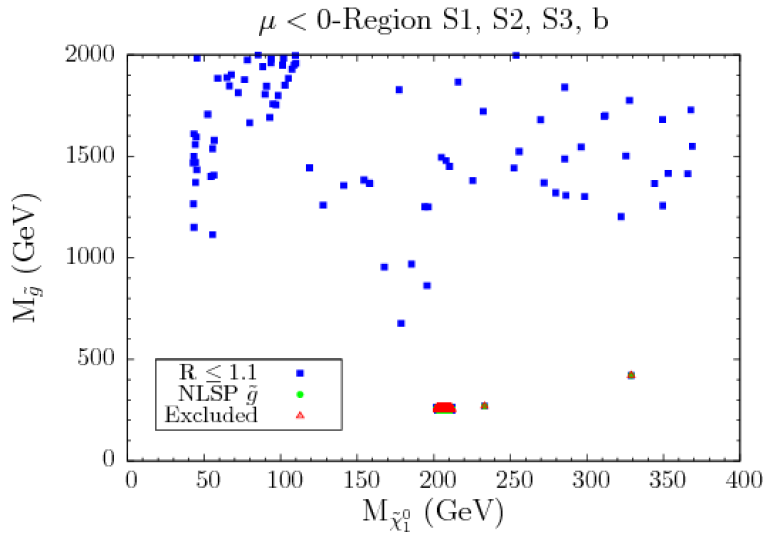}
\end{center}
\caption{$M_{\tilde{g}}$ vs. $M_{\tilde{\chi}_1^0}$ for models with Yukawa unification, NLSP gluino and those excluded models by 
ATLAS region S1, S2, S3, b in the 4-2-2 framework with $\mu<0$. Blue regions correspond to models with Yukawa unification ($R\leq 1.1$). Green regions are a subset with NLSP gluino, and models excluded by LHC data are in red color.} 
\label{nmumgmx}
\end{figure}

\begin{table}[tb]
\begin{center}
\begin{tabular}[t]{|c|c|c|c|c|c|}
  \hline
 $R\leq 1.1 \ \&$ NLSP $\tilde g$ & S1 & S2 & S3 & b & S1, S2, S3, b\\
  \hline
  excluded & 3800 & 3807 & 3385 & 3551 & 3807 \\
  \hline
 fraction & 99.8$\%$ & 100$\%$ & 88.9$\%$ & 93.3$\%$ & 100$\%$\\
  \hline 
\end{tabular}
\end{center}
\caption{Number of excluded 4-2-2 models with Yukawa unification ($R\leq 1.1$) and NLSP gluino for $\mu<0$. Also shown are the exclusion fraction by individual signal regions S1, S2, S3, b, and by combined S1, S2, S3, b.}
\label{4221}
\end{table}

The CMS analysis has stated less stringent constraints than ATLAS for low-energy supersymmetry search~\cite{nath,martin}, and so we utilize the requirements used by ATLAS in our studies below. In the updated analysis for multi-jets and missing energy, the events are classified into 3 regions ``S1'', ``S2'' and ``S3'', where S1, S2, S3 requires at least 2, 3, 4 jets respectively~\cite{atlasnew}. The second class of analysis requires at least one heavy flavor $b$-jet  candidate in final states~\cite{atlas2}, denoted by ``b'' in the following. To simulate similar data, we generate all supersymmetric $2\to 2$ events and include parton showering and hadronization using Pythia~\cite{pythia}, and then forward them to fast detector simulation PGS-4~\cite{pgs} to simulate the important detector effects. The $b$-tagging efficiency and mis-tagging rate in PGS-4 are based on the Technical Design Reports of ATLAS, and we use the default values in our analysis. We further follow the same ATLAS selection cuts for S1, S2, S3 and b.
The cut requirements are summarized in Table~\ref{cuts}, where $\Delta \phi(\vec{p}_T^{{\rm miss}}, j_{1,2,3})$ is the smallest azimuthal separation between the $\cancel{E}_T$ direction and the three leading jets, and $m_{eff}$ is the scalar sum of $\cancel{E}_T$ and the transverse momenta of the highest $p_T$ jets (up to two for region S1, three for region S2 and four for regions S3 and b respectively). The 95$\%$ C.L. upper limits on effective cross section (cross-section times acceptance) for non-Standard Model (SM) processes for signal region S1, S2, S3, b are also showed in Table~\ref{cuts}. Following Ref.~\cite{martin} we apply $\sigma\times {\rm acceptance}>\sigma_{{\rm exp}}$ as exclusion requirement for each model, where $\sigma$ is the relevant total cross-section and the acceptance is the ratio of signal events after and before selection cuts which reflects the effects of experimental efficiency.

\section{LHC C\lowercase{onstraints} \lowercase{on} NLSP G\lowercase{luino} \lowercase{and} N\lowercase{eutralino} D\lowercase{ark} M\lowercase{atter}}

\subsection{$t-b-\tau$ Yukawa Unification with $\mu<0$}
In Refs.~\cite{4221,4222}, the supersymmetric 4-2-2 models with $t-b-\tau$ Yukawa unification are studied for positive~\cite{4221} and negative~\cite{4222} signs of the MSSM parameter $\mu$. The $SU(2)_L$ gaugino mass $M_2$ was chosen to have the same sign as $\mu$ in order to remain consistent with the $(g-2)_\mu$ measurement. This is because the supersymmetric contribution to $(g-2)_\mu$ is proportional to $\mu M_2$. In this section we first consider the ATLAS constraints on 4-2-2 models with $\mu<0$. In this case, the finite threshold correction to the Yukawa coupling $y_b$ involving the gluino has the desired negative sign. Namely~\cite{btau},
\begin{eqnarray}
\delta y_b^{\rm SUSY-finite}&\approx&{g_3^2\over 12\pi^2}{\mu M_{\tilde{g}}\tan\beta \over M_b^2}+
{y_t^2\over 32\pi^2}{\mu A_t\tan\beta\over M_t^2},
\label{yb}
\end{eqnarray}
where $g_3$ is the strong gauge coupling, $A_t$ is the stop trilinear coupling, and $M_b\approx (M_{\tilde{b}_1}+M_{\tilde{b}_2})/2$, $M_t\approx (M_{\tilde{t}_2}+\mu)/2$.
For the desired Yukawa unification ($\approx 10\%$ or better), one obtains a wide range of acceptable gluino masses, namely $\mathcal{O}(200)\lesssim M_{\tilde{g}} \lesssim \mathcal{O}(1000)$ GeV as shown in Fig. 5 of Ref.~\cite{4222}. In particular, for relatively light gluinos, Yukawa unification is compatible with the gluino-bino co-annihilation mechanism and requires near-degenerate NLSP gluino and LSP neutralino masses.

\begin{figure}[tb]
\begin{center}
\includegraphics[scale=1,width=8cm]{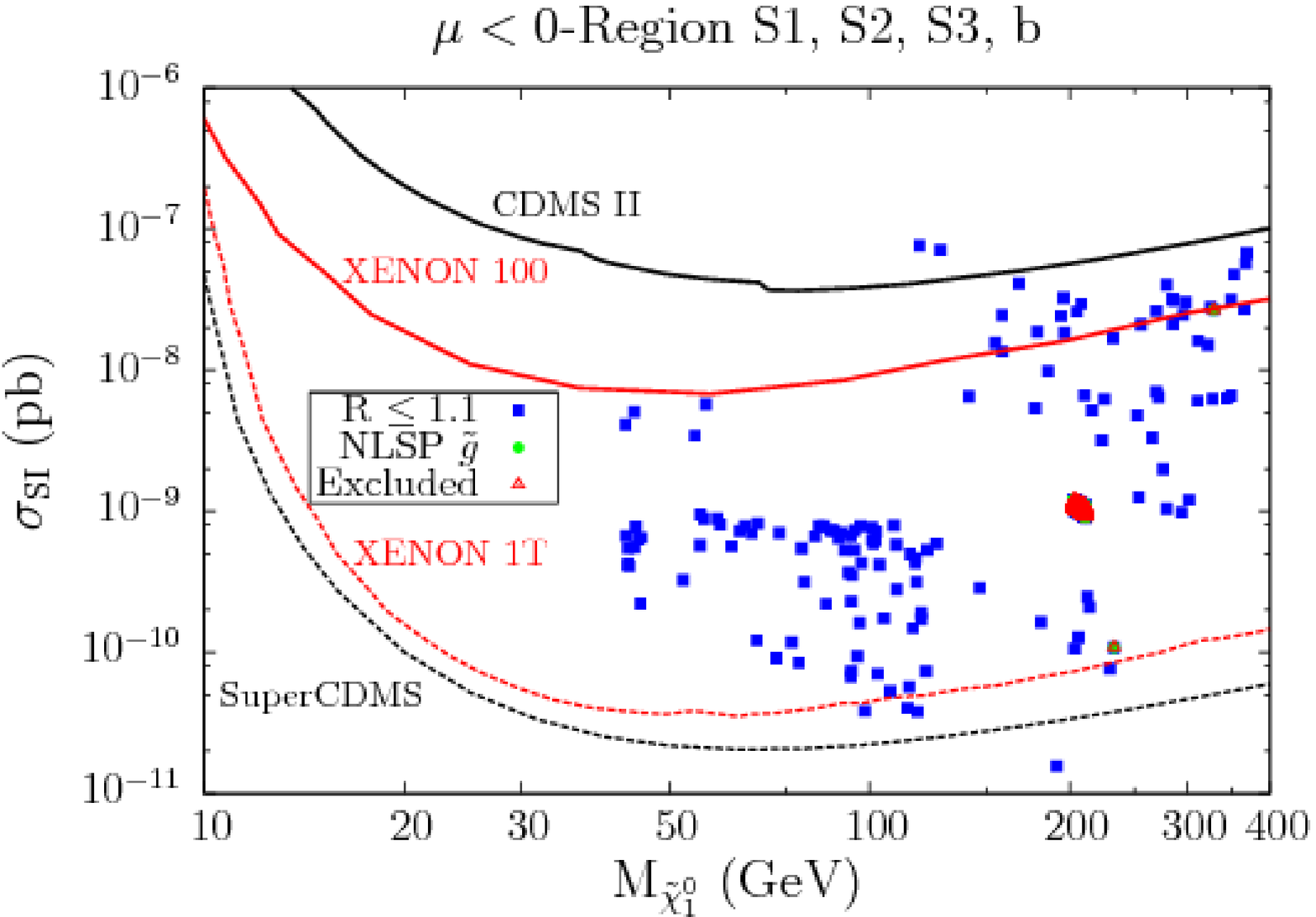}
\includegraphics[scale=1,width=8cm]{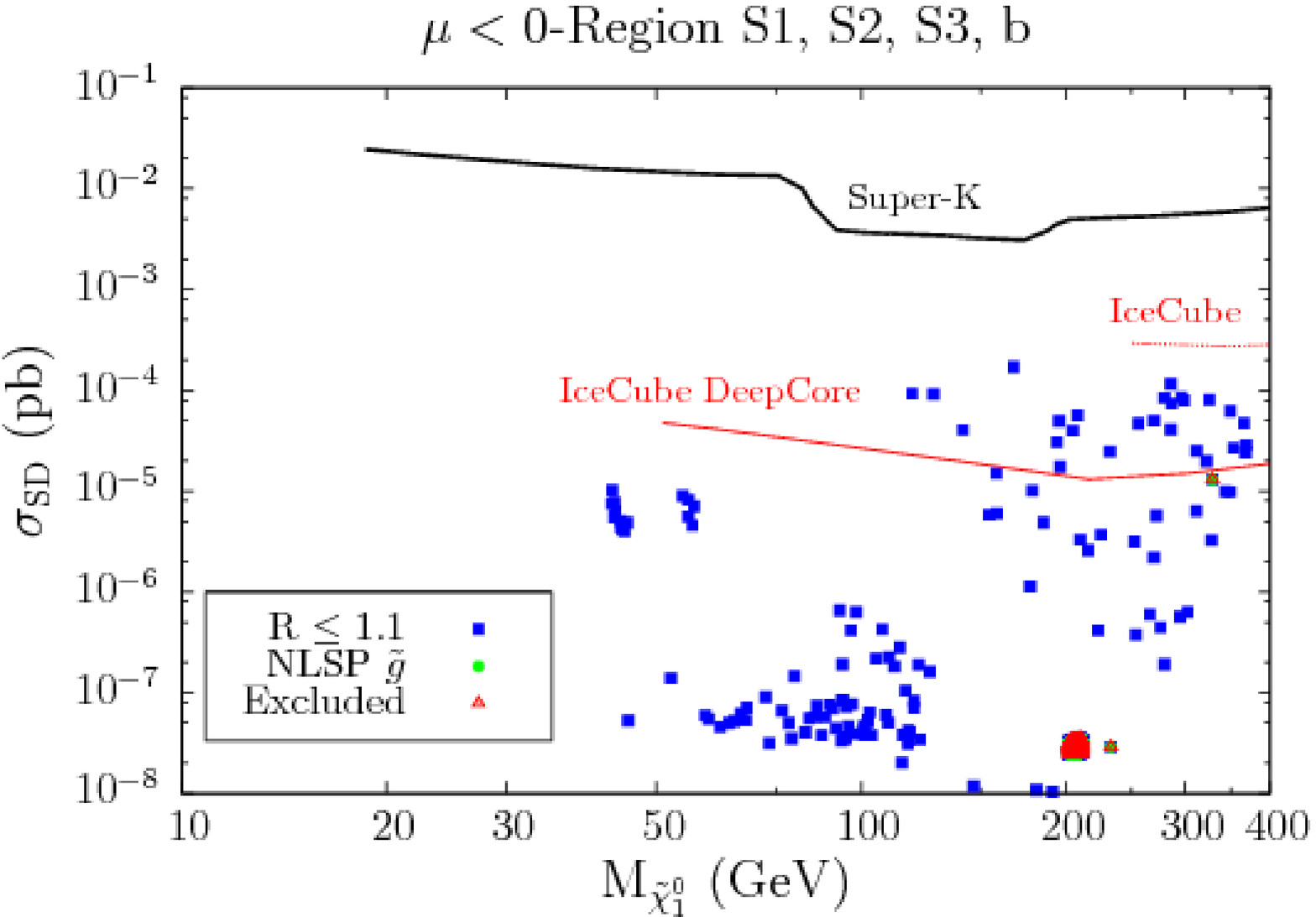}
\end{center}
\caption{$\sigma_{SI}$ (left panel) and $\sigma_{SD}$ (right panel) vs. $M_{\tilde{\chi}_1^0}$ in 4-2-2 models with Yukawa unification, NLSP gluino, and $\mu<0$. The excluded region is denoted in red. The current limits from CDMS-II, XENON100, SuperK and IceCube and future projected sensitivities from XENON1T, SuperCDMS and IceCube DeepCore are also shown.} 
\label{nmudd}
\end{figure}

\begin{figure}[tb]
\begin{center}
\includegraphics[scale=1,width=10cm]{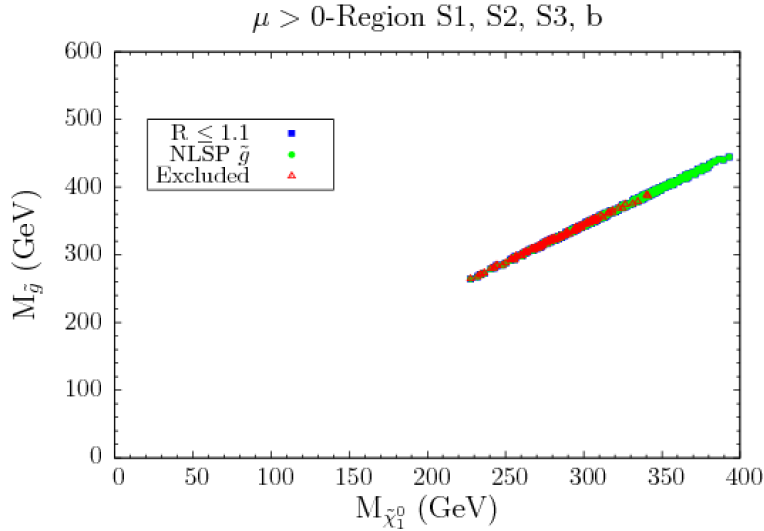}
\end{center}
\caption{$M_{\tilde{g}}$ vs. $M_{\tilde{\chi}_1^0}$ for models with Yukawa unification ($R\leq 1.1$, in blue), NLSP gluino (subset, in green) and those excluded by 
ATLAS region S1, S2, S3, b (in red), for $\mu>0$.} 
\label{pmumgmx}
\end{figure}

 To study the LHC constraints on this class of models, we generate about half a million models  by scanning the parameter space \cite{4222} and finally obtain 5420 models after applying the various experimental constraints listed in section \ref{sec2}. Out of these, 3945 models have acceptable Yukawa unification ($R \leq 1.1$), and in  3807  of these models gluino is the NLSP. The region in which the NLSP gluino and LSP neutralino are nearly mass degenerate corresponds to $250 \ \mathrm{GeV}\lesssim M_{\tilde{g}} \lesssim 300$ GeV. 
\begin{figure}[tb]
\begin{center}
\includegraphics[scale=1,width=8cm]{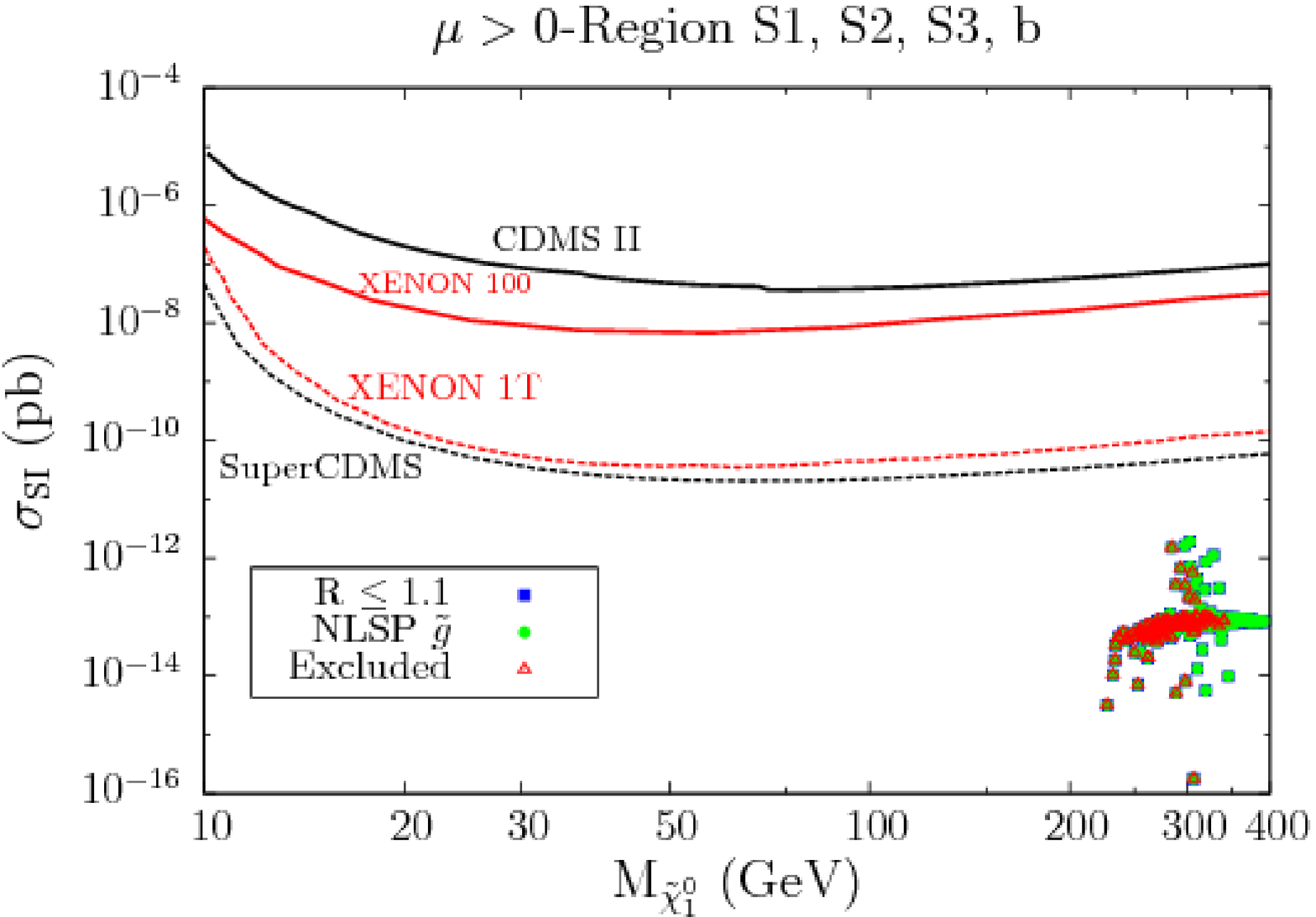}
\includegraphics[scale=1,width=8cm]{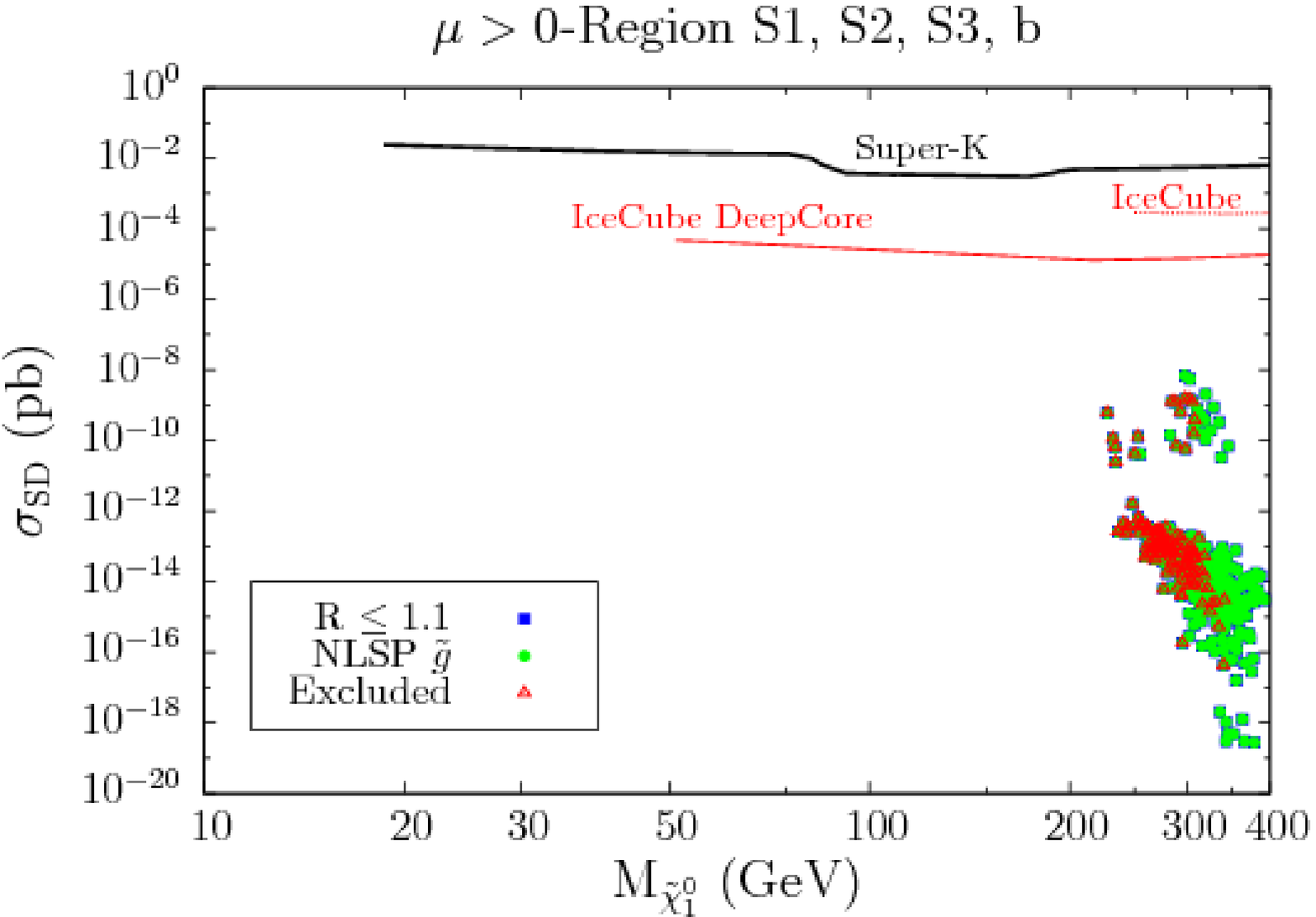}
\end{center}
\caption{$\sigma_{SI}$ (left panel) and $\sigma_{SD}$ (right panel) vs. $M_{\tilde{\chi}_1^0}$. Color scheme is same as in Fig.~\ref{nmudd}. The current limits from CDMS-II, XENON100, SuperK and IceCube and future anticipated bounds from XENON1T, SuperCDMS and IceCube DeepCore are also shown.} 
\label{pmudd}
\end{figure}

\begin{table}[th!]
\centering
\begin{tabular}{lccc}
\hline
\hline
                 & Point 1     & Point 2 & Point 3    \\
\hline
$m_{0}$  &  1511  &    10317  &  19639\\
$M_{1}$  &  -468.24  &    436.59  &  672.06\\
$M_{2}$  &  -826.2  &    719.35  &  1119.4\\
$M_{3}$  &  68.7  &    12.45  &  1.05\\
$\tan\beta$  &  47.5    &  49.66  &  50.93\\
$A_0$  &  -1680.23  &    -24285  &  -49722\\
sgn$(\mu)$  &  -1  &    +1  &  +1\\
$m_{Hu}$  &  505.5  &    3,550.53  &  7,964.78\\
$m_{Hd}$  &  1,029.83    &  10,288.17  &  16,115.48\\
\hline
$m_h$  &  114   &  125  &  126\\
$m_H$  &  445  &  6,307  &  6,631\\
$m_A$  &  442   &  6,266  &  6,588\\
$m_{H^{\pm}}$  &  454  &  6,308  &  6,632\\
\hline
$m_{\tilde{\chi}^0_{1,2}}$  &       202,     684   &      237,    737  &      390,   1204\\
$m_{\tilde{\chi}^0_{3,4}}$  &     1136,    1144    &    10231,   10231  &    20043,   20043\\
$m_{\tilde{\chi}^{\pm}_{1,2}}$  &       685,    1144    &   740,  10218  &     1208,  20037\\
$m_{\tilde{g}}$  &  258   &  276  &  447\\
\hline
$m_{\tilde{u}_{L,R}}$  &      1595,    1503   &     10323,   10155  &     19649,   19482\\
$m_{\tilde{t}_{1,2}}$  &       996,    1163    &      4291,    4712  &      6887,    7953\\
\hline
$m_{\tilde{d}_{L,R}}$  &      1597,    1515    &     10324,   10381  &     19649,   19728\\
$m_{\tilde{b}_{1,2}}$  &       971,    1172   &      4384,    4715  &      7717,    8379\\
\hline
$m_{\tilde{\nu}_{1}}$  &  1595    &  10222  &  19550\\
$m_{\tilde{\nu}_{3}}$  &  1416    &  7785  &  15082\\
\hline
$m_{\tilde{e}_{L,R}}$  &      1597,    1533   &     10221,   10526  &     19547,   19861\\
$m_{\tilde{\tau}_{1,2}}$  &      1119,    1421   &      4850,    7775  &      9338,   15025\\
\hline
$\sigma_{SI}({\rm pb})$  &  1.14$\times 10^{-9}$   &  4.48$\times 10^{-14}$  &  8.17$\times 10^{-14}$\\
$\sigma_{SD}({\rm pb})$  &  3.06$\times 10^{-8}$    &  2.60$\times 10^{-13}$  &  3.52$\times 10^{-15}$\\
$\Omega_{CDM}h^2$  &  0.11   &  0.10  &  0.09\\
$R$  &  1.04   &  1.08  &  1.04\\
\hline
$\sigma \times \mathrm{acc \ (S1) \ (pb)}$  &  0.133    &  0.073  &  0.012\\
$\sigma \times \mathrm{acc \ (S2)\ (pb)}$  &  0.158   &  0.048  &  0.018\\
$\sigma \times \mathrm{acc \ (S3)\ (pb)}$  &  0.091    &  0.03  &  0.006\\
$\sigma \times \mathrm{acc \ (b)\ (pb)}$  &  0.6    &  0.2  &  0\\
\hline
\hline
\end{tabular}
\caption{{LHC limits on three NLSP gluino benchmark points that satisfy all the experimental constraints described in Section \ref{sec2}. Various selection cuts from the four  signal regions, namely, S1, S2, S3 and b exclude point 1, whereas point 2 is excluded by the first two regions. Point 3 is allowed by all four signal regions.}
\label{table3}}
\end{table}

In Fig.~\ref{nmusigma} we show $\sigma\times$acceptance vs. $M_{\tilde{g}}$ for 4-2-2 models with Yukawa unification, using the ATLAS regions S1, S2, S3 and b. The near-degenerate NLSP-LSP points are also specified and actually overlap with the Yukawa unified points in the low gluino mass region. One can see that near-degenerate NLSP-LSP models with $M_{\tilde{g}}\lesssim 300$ GeV are essentially excluded. To display this perhaps more clearly, in Table~\ref{4221} we outline the number of excluded NLSP models with Yukawa unification and the excluded fraction for these models by individual signal regions S1, S2, S3, b and combined S1, S2, S3, b. Among the three regions S1, S2, S3 of multi-jets plus missing energy final states, region S3 is the weakest for constraining NLSP gluino because it requires four jets with $p_T>40$ GeV. However, the softest jet from a pair of NLSP gluinos more likely cannot have $p_T$ more than about 20 GeV. Therefore, a greater number of events do not pass the selection cuts of region S3 compared with region S1 and S2.
Furthermore, as we expect, region b with $b$-jets in the final states excludes a significant fraction of NLSP gluino models, although we employ the early LHC data with 35 pb$^{-1}$ integrated luminosity.  This is because the decay $\tilde{g}\to b\bar{b}\tilde{\chi}_1^0$ is dominant in most of the NLSP gluino region, which makes the NLSP gluino models more sensitive to multi-$b$ jets signature. 

In Fig.~\ref{nmumgmx} ($M_{\tilde{g}}-M_{\tilde{\chi}_1^0}$ plane), we display (in red color) the excluded models which have Yukawa unification and NLSP gluino. One can see that heavier gluinos with $M_{\tilde{g}}\gtrsim 500$ GeV are consistent with Yukawa unification, but being fairly massive, they survive the current LHC constraint. 

It is important to see the implications of LHC data on direct and indirect dark matter detection in this class of Yukawa unified models with NLSP gluino. 
In Fig.~\ref{nmudd} we display this by plotting the spin-independent and spin-dependent WIMP-nucleon scattering cross-section $\sigma_{SI}$ (left panel) and $\sigma_{SD}$ (right panel) vs. $M_{\tilde{\chi}_1^0}$. A significant region around $M_{\tilde{\chi}_1^0}\simeq 200$ GeV is excluded by LHC data, although it is allowed by CDMS-II, XENON100, SuperK and IceCube experiments. This excluded region will be
tested in the future by XENON 1T and SuperCDMS, but the region lies about three orders of magnitude below the expected IceCube DeepCore bound.





\subsection{$t-b-\tau$ Yukawa Unification with $\mu>0$}

With $\mu>0$, the gluino contribution to $\delta y_b^{\rm finite}$ is positive, so that the contribution from the chargino loop must be negative and sufficiently large in order to overcome this.
In this scenario lower gluino masses and larger values of $A_t$ and $M_b$ in Eq.~(\ref{yb}) are favored. All realistic NLSP gluino models compatible with the WMAP dark matter constraint give rise in this case to gluino masses in the range $220 \ \mathrm{GeV}\lesssim M_{\tilde{g}}\lesssim 400$ GeV, with the LSP neutralino closely degenerate in mass. Also, because of the large $A_t$ and $M_b$ values, in this scenario one of the stops is usually the lightest squark, with the sbottom relatively heavier than in $\mu<0$ case. Thus, the three-body decay $\tilde{g}\to b\bar{b}\tilde{\chi}_1^0$ through an off-shell sbottom is suppressed, so that the constraint from $b$-jets in the final states is less stringent. We start with about 1 million models and obtain 17942 models which survive the low-energy experiments. Out of these, about 400 models display acceptable Yukawa unification ($R\leq 1.1$) and contain NLSP gluino. 
Note that the constraint from $(g-2)_\mu$ is ignored in generating these models~\cite{4222}.

After applying the ATLAS selection cuts listed in Table~\ref{cuts} a significant number of models are excluded. We show this in Fig.~\ref{pmumgmx} in the $M_{\tilde{g}}-M_{\tilde{\chi}_1^0}$ plane. The subset of models with NLSP gluino which overlaps with Yukawa unification is also specified. 
NLSP gluino masses below about $250-300$ GeV are essentially excluded. 
In Fig.~\ref{pmudd} we display the spin-independent and spin-dependent WIMP-nucleon scattering cross-section $\sigma_{SI}$ (left panel) and $\sigma_{SD}$ (right panel) vs. $M_{\tilde{\chi}_1^0}$. One can see a significant region around $M_{\tilde{\chi}_1^0}\simeq 250$ GeV is excluded by the LHC data although it is allowed by CDMS-II, XENON100, SuperK and IceCube experiments. Indeed, some parts of the excluded parameter space lie beyond the reach of future experiments such as XENON 1T, SuperCDMS and IceCube DeepCore.

Finally, in Table \ref{table3} we present three characteristic benchmark points with NLSP gluino, dark matter neutralino and very acceptable t-b-$\tau$ Yukawa unification. Points 1 and 2, with gluino masses close to 300 GeV are excluded by the selection cuts listed in the table. However, point 3 with NLSP gluino mass close to 450 GeV is compatible with the data.





\section{S\lowercase{ummary}}
Inspired by the recent LHC search of final states containing jets and/or b-jet and missing transverse momentum, corresponding to an 
integrated luminosity of 165 pb$^{-1}$, we have explored its ramifications 
for supersymmetric $SU(4)_c\times SU(2)_L\times SU(2)_R$ models which display 
t-b-$\tau$ Yukawa unification at 10\% level or better, contain NLSP 
gluino, and possess LSP neutralino dark matter. The NLSP gluino 
primarily decays into the LSP neutralino and a gluon or quark-antiquark 
pair, thus allowing us to exploit this LHC data.
For $\mu<0$ we generate about 4000 models, from an initial half a million
models, which satisfy the above criteria of Yukawa unification, NLSP 
gluino, and neutralino dark matter, after imposing constraints from all 
previous experiments (except LHC). The corresponding number of models 
for $\mu>0$ is around 400. We next show that for closely mass 
degenerate NLSP gluino and LSP neutralino, models with NLSP glunio 
masses below 300 GeV or so are largely excluded by the LHC 
data. The LHC implications for spin-dependent and spin-independent LSP 
neutralino-nucleon cross sections are also explored. Regions of the 
parameter space, some lying well below the much anticipated future 
bounds from IceCube DeepCore and Xenon 1T and SuperCDMS, are already 
excluded by utilizing the LHC data.

\subsection*{Acknowledgment}
We would like to thank Gregg Peim, Bin He, Ilia Gogoladze, Rizwan Khalid and Shabbar Raza for useful discussions. This work is supported by the DOE under grant No. DE-FG02-91ER40626.



\end{document}